# Overview: Main Fundamentals for Steganography

Zaidoon Kh. AL-Ani, A.A.Zaidan, B.B.Zaidan and Hamdan.O.Alanazi

***Abstract*— Th**e rapid development of multimedia and internet allows for wide distribution of digital media data. It becomes much easier to edit, modify and duplicate digital information .Besides that, digital documents are also easy to copy and distribute, therefore it will be faced by many threats. It is a big security and privacy issue, it become necessary to find appropriate protection because of the significance, accuracy and sensitivity of the information. Steganography considers one of the techniques which used to protect the important information. The main goals for this paper, to recognize the researchers for the main fundamentals of steganography.  In this paper provides a general overview of the following subject areas: Steganography types, General Steganography system, Characterization of Steganography Systems and Classification of Steganography Techniques.

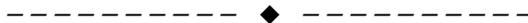

## 1. INTRODUCTION

Steganography is the art of hiding and transmitting data through apparently innocuous carriers in an effort to conceal the existence of the data, the word Steganography literally means covered or hiding writing as derived from Greek. Steganography has its place in security [1]. It is not intended to replace cryptography but supplement it. Hiding a message with Steganography methods reduces the chance of a message being detected. If the message is also encrypted then it provides another layer of protection. Therefore, some Steganographic methods combine traditional Cryptography with Steganography; the sender encrypts the secret message prior to the overall communication process, as it is more difficult for an attacker to detect embedded cipher text in a cover [2].   It has been used through the ages by ordinary people, spies, rulers, government, and armies .There are many stories about Steganography.

- *Zaidoon Kh. AL-Ani - Master Student on the faculty of Information and Communication Technology at International Islamic University Malaysia, Kuala Lumpur, Malaysia.*
- *A. A. Zaidan – PhD Candidate on the Department of Electrical & Computer Engineering , Faculty of Engineering , Multimedia University , Cyberjaya, Malaysia*
- *B. B. Zaidan – PhD Candidate on the Department of Electrical & Computer Engineering / Faculty of Engineering, Multimedia University, Cyberjaya, Malaysia*
- *Hamdan.O.Alanazi – Master Student, Department of Computer System & Technology, University Malaya, Kuala Lumpur, Malaysia.*

For example ancient Greece used methods for hiding messages such as hiding In the field of Steganography, some terminology has developed. The adjectives 'cover', 'embedded', and 'stego' were defined at the information hiding workshop held in Cambridge, England [5].   The term "cover" refers to description of the original, innocent massage, data, audio, video, and so on. Steganography is not a new science; it dates back to ancient times [7].
it in the belly of a share (a kind of rabbits), using invisible ink and pigeons [5]. Another ingenious method was to shave the head of a messenger and tattoo a message or image on the messenger head. After allowing his hair to grow, the message would be undetected until the head was shaved again. While the Egyptian used illustrations to conceal message. Hidden information in the cover data is known as the "embedded" data and information hiding is a general term encompassing many sub disciplines, is a term around a wide range of problems beyond that of embedding message in content [5]. The term hiding here can refer to either making the information undetectable or keeping the existence of the information secret [4],[5]. Information hiding is a technique of hiding secret using redundant cover data such as images, audios, movies, documents, etc [9]. This technique has recently become important in a number of application areas. For example, digital video, audio, and images are increasingly embedded with imperceptible marks, which may contain hidden signatures or watermarks that help to prevent unauthorized copy. It is a performance that inserts secret messages into a cover file, so that the existence of the messages is not apparent [7]. Research in information hiding has tremendous increased during the past decade with commercial interests driving the field [5],[7],[8].

Although the art of concealment "hidden information" as old as the history, but the emergence of computer and the evolution of sciences and techniques breathe life again in this art with the use of new ideas, techniques, drawing on the computer characteristics in the way representation of the data, well-known computer representation of all data including ( Multimedia) is binary these representations are often the digital levels and areas and change values-aware of slight not aware or felt by Means sensual of human such as hearing, sight, the advantage use of these properties to hide data in multimedia by replace the values of these sites to the values of data to be hidden, taking into account the acceptable limits for the changeover, and not exceeded to prevent degradation media container with a change becomes aware and felt by human [21]. It should be noted here that although the art of hidden information come in the beginning of the computer and its techniques However, the seriousness of the work in the stenography as a stand-alone science started in 1995. In this paper will address on the traditional and popular methods of hidden data within the multimedia computer also we will review of literature, in fact we will review some papers for the work has been done before, the authors have analysis these papers and shows them opinion of the main fundamentals for Steganography[14].

## 2. STEGANOGRAPHY TYPES

As it is known there is much communication between people and organizations through the use of the phone, the fax, computer communications, radio, and of course all of these communication should be secure. There are basically three Steganography types[14]:-

- Pure Steganography.
- Secret key Steganography.
- Public key Steganography

**2.1 Pure Steganography**

Pure Steganography is a Steganography system that doesn't require prior exchange of some secret information before sending message; therefore, no information is required to start the communication process: the security of the system thus depends entirely on its secrecy [14].

The pure Steganography can be defined as the quadruple (C, M, D, and E) where:
C: the set of possible covers.
M: the set of secret massage with $|C| \geq |M|$.
E: C×M→C the embedding function.

D: C→M of the extraction function with the property that
D (E(c,m))=m for all m $\in$ M and c $\in$ C

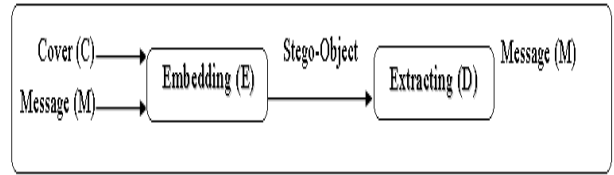

Fig 1. Pure Steganography

In most applications, pure Steganography is preferred, since no stego-key must be shared between the communication partners, although a pure Steganography protocols don't provide any security if an attacker knows the embedding method

**2.2 Secret Key Steganography**

A secret key Steganography system is similar to a symmetric cipher, where the sender chooses a cover and embeds the secret message into the cover using a secret key. If the secret key used in the embedding process is known to the receiver, he can reverse the process and extract the secret message [14].

Anyone who doesn't know the secret key should not be able to obtain evidence of the encoded information.

The secret key Steganography can be defined as the quintuple (C, M, K, DK, EK) where:
C: the set of possible covers.
M: the set of secret message.
K: the set of secret keys.
Ek: C×M×K→C
With the property that DK (EK(c,m,k),k)=m for all m $\in$ M, c $\in$ C and k $\in$ K

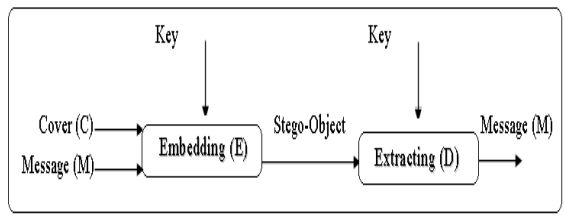

Fig 2.Secret Key Steganography.

**2.3 Public Key Steganography**

Public key Steganography does not depend on the exchange of a secret key. It requires two keys, one of them private (secret) and the other public: the public key is stored in a public database, whereas the public key is used in the embedding process. The secret key is used to reconstruct the secret message [14].

One way to build a public key Steganography system is to use a public key crypto system. The sender and the receiver can exchange public keys of some public key cryptography algorithm before imprisonment. Public key Steganography utilizes the fact that the decoding function in a Steganography system can be applied to any cover, whether or not it already contains a secret message [9].

The public key Steganography relies on the fact that encrypted information is random enough to hide in plain sight. The sender encrypts the information with the receiver's public key to obtain a random-looking massage and embeds it in a channel known to the receiver, thereby replacing some of the natural randomness with which every communication process is accompanied. Assume that both the cryptographic algorithms and the embedding functions are publicly known [8],[9].

The receiver who cannot decide a priori if secret information is transmitted in a specific cover will suspect the arrival of message and will simply try to extract and decrypt it using his private key. If the cover actually contained information, the decryption information is the sender's message [10].

## 3. GENERAL STEGANOGRAPHY SYSTEM

A general Steganography system is shown in Fig 3. It is assumed that the sender wishes to send via Steganographic transmission, a message to a receiver. The sender starts with a cover message, which is an input to the stego-system, in which the embedded message will be hidden. The hidden message is called the embedded message. A Steganographic algorithm combines the cover massage with the embedded message, which is something to be hidden in the cover .The algorithm may, or may not, use a Steganographic key (stego key), which is additional secret data that may be needed in the hidden process. The same key (or related one) is usually needed to extract the embedded massage again. The output of the Steganographic algorithm is the stego message. The cover massage and stego message must be of the same data type, but the embedded message may be of another data type. The receiver reverses the embedding process to extract the embedded message [13],[15],[16],[17],[18].

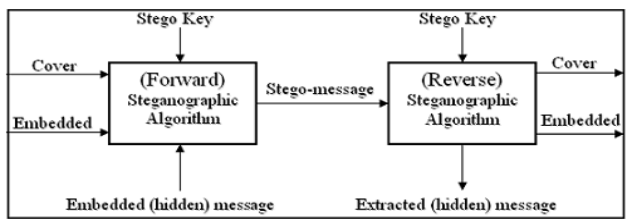

Fig 3. General Steganography System.

## 4. CHARACTERIZATION OF STEGANOGRAPHY SYSTEMS

Steganographic techniques embed a message inside a cover. Various features characterize the strength and weaknesses of the methods. The relative importance of each feature depends on the application [19],[14].

### 4.1 Capacity

The notion of capacity in data hiding indicates the total number of bits hidden and successfully recovered by the Stego system [20],[14].

### 4.2 Robustness

Robustness refers to the ability of the embedded data to remain intact if the stego-system undergoes transformation, such as linear and non-linear filtering; addition of random noise; and scaling, rotation, and loose compression [21].

### 4.3 Undetectable

The embedded algorithm is undetectable if the image with the embedded message is consistent with a model of the source from which images are drawn. For example, if a Steganography method uses the noise component of digital images to embed a secret message, it should do so while not making statistical changes to the noise in the carrier. Undetectability is directly affected by the size of the secret message and the format of the content of the cover image [14].

### 4.4 Invisibility (Perceptual Transparency)

This concept is based on the properties of the human visual system or the human audio system. The embedded information is imperceptible if an average human subject is unable to distinguish between carriers that do contain hidden information and those that do not. It is important that the embedding occurs without a significant degradation or loss of perceptual quality of the cover [14],[21].

### 4.5 Security

It is said that the embedded algorithm is secure if the embedded information is not subject to removal after being discovered by the attacker and it depends on the total information about the embedded algorithm and secret key [11],[17],[18],[20].

# 5. CLASSIFICATION OF STEGANOGRAPHY TECHNIQUES

There are several approaches in classifying Steganographic systems. One could categorize them according to the type of covers used for secret communication or according to the cover modifications applied in the embedding process. The second approach will be followed in this section, and the Steganographic methods are grouped in six categories, although in some cases an exact classification is not possible. Fig 4. Presents the Steganography classification [8],[9],[10],[15].

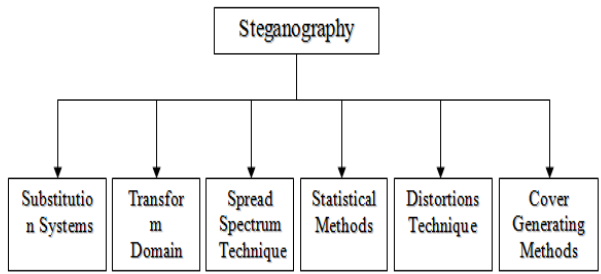

Fig 4. Steganography Classification.

## 5.1 Substitution Systems

Basic substitution systems try to encode secret information by substituting insignificant parts of the cover by secret message bits. The receiver can extract the information if he has knowledge of the positions where secret information has been embedded. Since only minor modifications are made in the embedding process, the sender assumes that they will not be noticed by an attacker. It consists of several techniques that will be discussed in more detail, in the following subsection:[9],[10],[11],[15].

### 5.1.1 Least Significant Bit Substitution

The embedding process consists of choosing a subset $\{j_1 \ldots j_{l(m)}\}$ of cover elements and performing the substitution operation $c_{j_i} \leftrightarrow m_i$ on them, which exchange the LSB of $c_{j_i}$ by $m_i$ ($m_i$ can be either 1 or 0).In the extraction process, the LSB of the selected cover-element is extracted and lined up to reconstruct the secret message .

### 5.1.2 Pseudorandom Permutation

If all cover bits are accessed in the embedding process, the cover is a random access cover, and the secret message bits can be distributed randomly over the whole cover. This technique further increases the complexity for the attacker, since it is not guaranteed that the subsequent message bits are embedded in the same order .

### 5.1.3 Image Downgrading and Cover Channels

Image downgrading is a special case of a substitution system in which image acts both as a secret message and a cover. Given cover-image and secret image of equal dimensions, the sender exchanges the four least significant bits of the cover grayscale (or color) values with the four most significant bits of the secret image. The receiver extracts the four least significant bits out of the stego-image, thereby gaining access to the most significant bits of the stego-image. Whereas the degradation of the cover is not visually noticeable in many cases, four bits are sufficient to transmit a rough approximation of the secret image

### 5.1.4 Cover Regions and Parity Bits

Any nonempty subset of $\{c_1,\ldots\ldots, c_{I(c)}\}$ is called a cover-region. By dividing the cover into several disjoint regions, it is possible to store one bit of information in a whole cover-region rather than in a single element. A parity bit of a region I can be calculated by:

$$B(I) = \sum_{j \in I} LSB(c_j) \bmod 2$$

### 5.1.5 Palette-Based Image

There are two ways to encode information in a palette-based image; either the palette or the image data can be manipulated. The LSB of the color vectors could be used for information transfer, just like the substitution methods presented. Alternatively, since the palette does not need to be sorted in any way, information can be encoded in the way the colors are stored in the palette. For N colors since there are N! Different ways to sort the palette, there is enough capacity to encode a small message. However, all methods which use the order of a palette to store information are not robust, since an attacker can simply sort the entries in a different way and destroy the secret message.

### 5.1.6 Quantization and Dithering

Dithering and quantization to digital image can be used for embedding secret information. Some Steganographic systems operate on quantized images. The difference $e_i$ between adjacent pixels $x_i$ and $x_i+1$ is calculated and fed into a quantize φ which outputs a discrete approximation ΔI of the different signal ($x_i - x_i+1$). Thus in each quantization step a quantization error is introduced. In order to store the ith message bit in the cover-signal, the quantized difference signal ΔI is computed. If according to the secret table ΔI does not match the secret bit to be encoded, ΔI is replaced by the nearest ΔI where the associated bit equals the secret message bit. The resulting value ΔI is those fed into the entropy coder. At the receiver side, the message is decoded according to the difference signal ΔI and the stego-key.

### 5.1.7 Information Hiding in Binary Image

Binary image contains redundancies in the way black and white pixels are distributed. Although the implementation of a simple substitution scheme is possible, these systems are highly susceptible to transmission errors and therefore are not robust.

### 5.1.8 Unused or Reversed Space in Computer Systems

Taking advantage of an unused or reversal space to hold covert information provides a means of hiding information without perceptually degrading the carrier. For example, the way operation systems store files typically results in an unused space that appears to be allocated to a file. Another method of hiding information in file system is to create hidden partitions. These partitions are not seen if the system is started normally.

## 5.2 Transform Domain Techniques

It has been seen that the substitution and modification techniques are easy ways to embed information, but they are highly vulnerable to even small modification. An attacker can simply apply signal processing techniques in order to destroy the secret information. In many cases even the small changes resulting out of loose compression systems yield total information loss. It has been noted in the development of Steganographic systems that embedding information in the frequency domain of a signal can be much more robust than embedding rules operating in the time domain. Most robust Steganographic systems known today actually operate in some sort of transform domain. Transformation domain methods hide message in a significant area of the cover image which makes them more robust to attack, such as adding noise, compression, cropping some image processing. However, whereas they are more robust to various kinds of signal processing, they remain imperceptible to the human sensory system. Many transform domain variations exist. One method is to use the Discrete Cosine Transformation (DCT) as a vehicle to embed information in image. Another method would be the use of wavelet transforms .Transforms embedding embeds a message by modification (selected) transform (e.g., frequency) coefficient of the cover message. Ideally, transform embedding has an effect on the spatial domain to apportion the hidden information through different order bits in a manner that is robust, but yet hard to detect. Since an attack, such as image processing, usually affects a certain band of transform coefficient, the remaining coefficient would remain largely intact. Hence, transform embedding is, in general, more robust than other embedding methods [13].

## 5.3 Spread Spectrum (SS) Techniques

Spread spectrum techniques are defined as "Means of transmission in which the signal occupies a bandwidth in excess of the minimum necessary to send the information". The band spread is accomplished by means of a code which is independent of the data, and a synchronized reception with the code at the receiver is used for dispreading and subsequent data recovery. Although the power of the signal to be transmitted can be large, the signal–to–noise ratio in every frequency band will be small, even if parts of the signal could be removed in several frequency bands, enough information should be present in the other bands to recover the signal. This situation is very similar to a Steganography system which tries to spread a secret message over a cover in order to make it impossible to perceive. Since spread signals tend to be difficult to remove, embedding methods based on SS should provide a considerable level of robustness .In information hiding, two special variants of spread spectrum techniques are generally used: direct sequence, and frequency–hopping scheme. In direct-sequence scheme, the secret signal is spread by a constant called ship rate, modulated with a pseudorandom signal and added to the cover. On the other hand, in the frequency–hopping schemes the frequency of the carrier signal is altered in a way that it hops rapidly from one frequency to another. SS are widely used in the context of watermarking [13],[14],[18],[21].

## 5.4 Distortion Techniques

In contrast to substitution systems, distortion requires the knowledge of the original cover in the decoding process.The sender applies a sequence of modifications to the cover in order to get a stego-system [19].A sequence of modification is chosen in such a way that it corresponds to a specific secret message to be transmitted. The receiver measures the difference in the original cover in order to reconstruct the sequence of modification applied by the sender, which corresponds to the secret message [14],[19]. An early approach to hiding information is in text. Most text-based hiding methods are of distortion type (i.e,the arrangement of words or the layout of a docement may reveal information). One technique is by modulating the positions of line and words, which will be detailed in the next subsection. Adding spaces and "invisible" characters to text provides a method to pass hidden information HTML files are good candidates for including extra spaces, tabs, and linebreaks. Web browsers ignore these "extra" spaces and lines and they go unnoticed until the sourse of the web pag is revealed [14],[19].

### 5.5 Cover Generation Techniques

In contrast to all embedding methods presented above, when secret information is added to a specific cover by applying an embedding algorithm, some Steganographic applications generate a digital object only for the purpose of being a cover for secret communication [14],[19].

Table 1
Weaknesses of Steganography Techniques

| Steganography Techniques | Weakness |
| --- | --- |
| Substitution Systems | Low robustness: filtering, lossy compression attacks, format file dependand. |
| Transform Domain Techniques | An attacker can simply apply signal processing techniques in order to destroy the secret information. In many cases even the small changes resulting out of loose compression systems yield total information loss. |
| Spread Spectrum (SS) Techniques | There are increases in the complexity, higher costs and more stringent timing requirements.<br>a) Direct-Sequence Scheme: The circuitry required to produce the spectrum is complex, it requires a large bandwidth channel with relatively small phase distortions and requires a long acquisition time since the PN codes are long.<br>b) Frequency–Hopping Scheme: Weakness with both slow and fast hopping. With slow hopping, coherent data detection is possible, but data can be lost if a single frequency hop channel is jammed. To overcome this, it is necessary to use error correcting codes. Fast hopping disposes of the need for error codes since one bit of data is spread over a number of hops. However, fast hopping has the disadvantage that due to phase discontinuities, coherent data detection is not possible. |
| Distortion Techniques | In many applications, such systems are not useful, since the receiver must have access to the original cover. It is weakness point. So if the attacker also has access to them, he/she can easily detect the cover modification and has evidence for a secret communication. If the embedding and extraction functions are public and do not depend on a stego-key, it is also possible for the attacker to reconstruct secret message entirely. |
| Cover Generation Techniques | They have heavy and complexity process for algorithems compersion with other techniques. This point due to dealy time for finished ( hiding or extract) process operation .Example: Automated Generation of English Text. Use a large dictionary of words categorised by different types, and a style source which describes how words of different types can be used to form a meaningful sentence. Transform message bits into sentences by selecting words out of the dictionary which conforms to a sentence structure given in the style source. |

### 5.6 Statistical Steganography

Statistical Steganography techniques utilize the existence of "1-bits" Steganography schemes, which embed one bit of information in a digital carrier. This is done by modifying the cover in such a way that some statistical characteristics change significantly if a "1" is transmitted. Otherwise, the cover is left UN changed. So the receiver must be able to distinguish unmodified covers from modified ones. A cover is divided into l (m) disjoint blocks B1...B l (m). A secret bit, mi is inserted into the ith block by placing "1" in to Bi if mi=1.Otherwise, the block is not changed in the embedding process [14],[15].

## 6. CONCLUSION

In this paper a overview for the Main Fundamentals of Steganography were presented in to five categories, firstly types of steganography: There are basically three Steganography types, Pure Steganography, Secret key Steganography and Public key Steganography A discussed about the each type with summarized. Secondly, General Steganography System: in this part we explain the Steganographic transmission system between sender and receiver. Thirdly, Characterization of Steganography Systems: overview various features characterize the strength and weaknesses of the methods of steganography. Finally, Classification of Steganography Techniques: Steganographic methods are grouped in six categories, although in some cases an exact classification is not possible.

## ACKNOWLEDGEMENT

Thanks in advance for the entire worker in this project, and the people who support in any way, also I want to thank UUM for the support which came from them.

**Zaidoon Kh. AL-Ani: H**as gained his bachelor Degree in computer science from University of Baghdad in 2003. Presently, he is conducting his master degree in Information and Communication Technology at International Islamic University Malaysia. The field of interest is steganography. Projects have been done which are titled as "Data security using hiding data" and Evaluation of Steganography for Arabic text".

**Aos Alaa Zaidan**: He obtained his 1st Class Bachelor degree in Computer Engineering from university of Technology / Baghdad followed by master in data communication and computer network from University of Malaya. He led or member for many funded research projects and He has published more than 55 papers at various international and national conferences and journals, His interest area are Information security (Steganography and Digital watermarking), Network Security (Encryption Methods) , Image Processing (Skin Detector), Pattern Recognition , Machine Learning (Neural Network, Fuzzy Logic and Bayesian) Methods and Text Mining and Video Mining. .Currently, he is PhD Candidate on the Department of Electrical & Computer Engineering / Faculty of Engineering / Multimedia University / Cyberjaya, Malaysia. He is members IAENG, CSTA, WASET, and IACSIT. He is reviewer in the (IJSIS, IJCSN, IJCSE and IJCIIS).

**Bilal Bahaa Zaidan:** He obtained his bachelor degree in Mathematics and Computer Application from Saddam University/Baghdad followed by master in data communication and computer network from University of Malaya. He led or member for many funded research projects and He has published more than 55 papers at various international and national conferences and journals, His interest area are Information security (Steganography and Digital watermarking), Network Security (Encryption Methods) , Image Processing (Skin Detector), Pattern Recognition , Machine Learning (Neural Network, Fuzzy Logic and Bayesian) Methods and Text Mining and Video Mining. .Currently, he is PhD Candidate on the Department of Electrical & Computer Engineering / Faculty of Engineering / Multimedia University / Cyberjaya, Malaysia. He is members IAENG, CSTA, WASET, and IACSIT. He is reviewer in the (IJSIS, IJCSN, IJCSE and IJCIIS).

**Hamdan Al-Anazi**: has obtained his bachelor dgree from "King Suad University", Riyadh, Saudi Arabia. He worked as a lecturer at Health College in the Ministry of Health in Saudi Arabia, then he worked as a lecturer at King Saud University in the computer department. Currently he is Master candidate at faculty of Computer Science & Information Technology at University of Malaya in Kuala Lumpur, Malaysia. His research interest on Information Security, cryptography, steganography and digital watermarking, He has contributed to many papers some of them still under reviewer.